\documentclass[12pt,preprint]{aastex}

\begin{document}

\title{A new method of estimating the mass-to-light ratio of the Ursa Minor dwarf
spheroidal galaxy }

\author{M. \'Angeles G\'omez-Flechoso\footnote{Present address: Universidad
Europea de Madrid, E-28670 Villaviciosa de Od\'on, Madrid, Spain}}
\affil{Geneva Observatory, Ch. des Maillettes 51, CH-1290 Sauverny,
Switzerland}

\author{D. Mart\'\i nez-Delgado}
\affil{Instituto de Astrof\'\i sica de Canarias, E-38205 La Laguna,
Tenerife, Canary Islands, Spain} \authoremail{ddelgado@ll.iac.es}

\begin{abstract}
Dwarf satellite galaxies undergo strong tidal forces produced by the main
galaxy potential. These forces disturb the satellite, producing asymmetries
in its stellar distribution, tidal tail formation, and modifications of
the velocity dispersions profiles.
Most of these features are observed in the Ursa Minor (UMi) dwarf spheroidal galaxy, which is
one of the closest satellites of the Milky Way. These features show that UMi is been
tidally disrupted and probably not in virial equilibrium. The high velocity dispersion
of UMi could also be a reflection of this tidal disruption and not the signature of the large
dark matter content that would be deduced if virial equilibrium is assumed.
In order to avoid the uncertainty produced when virial equilibrium is assumed
in systems in strong tidal fields, we present a new method of calculating the
mass-to-luminosity ratio of disrupted dwarf galaxies. This method is based on numerical
simulations and only takes into account the shape of the dwarf density profile and
the tidal tail brightness, but  does not depend on the kinematics of the dwarf.
Applying this method to UMi, we obtain a mass-to-luminosity relation of 12, which
is lower than the value obtained assuming virial equilibrium ($M/L=60$). In addition,
if UMi has a large dark-matter content, it will be impossible to reproduce a tidal
tail as luminous as the one observed.

\end{abstract}

\section{Introduction }
\label{intro}

The simplest galactic
systems, the dwarf spheroidal galaxies (dSphs), reveal high velocity
dispersions that imply the highest known mass-to-light ratios ($M/L$)
(see Mateo 1998 for a review). These
large velocity dispersions are commonly interpreted in terms of the galaxies being embedded
in a massive dark-matter halo. Alternative explanations are
these mass-to-light ratios may be inflated owing to tidal effects (Hodge \& Michie 1969;
G\'omez-Flechoso, Fux, \& Martinet 1999), or that these systems are actually
long-lived tidal remnants oriented close to the line of sight (Kroupa 1997).

Ursa Minor (UMi) is one of the closest satellites of the Milky Way ($d = 69$ kpc;
Carrera, Aparicio, \& Mart\'\i nez-Delgado 2002) and, of these, shows, together with Draco,
the highest observed velocity
dispersion ($M/L=60$; Mateo 1998). The discovery of a tidal extension in UMi
(Mart\'\i nez-Delgado et al. 2001)  indicates that this
galaxy is at an advanced stage of complete tidal disruption (see also Palma
et al. 2002).  In addition, the presence of lumpiness and asymmetry in the stellar
distribution of UMi along its major axis (Olszewski \& Aaronson 1985; Irwin \& Hatzidimitriou
1995; Kleyna et al. 1998; Mart\'\i nez-Delgado et al. 2001) could be reminiscent of
tidally disrupted satellites (Kroupa 1997). The overall shape of the outer contours
appears to be clearly S-shaped (Palma et al 2002; Mart\'\i nez-Delgado et al. 2002, submitted),
as is expected for a tidally disrupted system (i.e., the globular cluster Pal 5; see
Odenkirchen et al. 2001). These observational features suggest that UMi may not be in
virial equilibrium, supporting the idea of a possible tidal origin for  UMi's
observed high radial velocity dispersion. This fuels the debate about the validity of
the methodology used to infer the high $M/L$ in dSphs and therefore the presence of large
quantities of dark matter in these galaxies.

In this paper, we introduce a new method of estimating the mass of satellite dSph galaxies
that are subjected to tidal forces. This method does not involve assumptions about the
internal dynamics of the satellite can and therefore
 be used for satellites that are out of virial equilibrium. Using this method, we
have analyzed the case of UMi and obtained an new value for the $M/L$ ratio of this galaxy.

\section{OBSERVATIONS AND TIDAL TAIL FORMATION MODEL}
\label{observations}

Observations, data reduction, and photometry of the UMi dSph galaxy used in this paper
are described in detail in Mart\'\i nez-Delgado et al. (2001) and Carrera et al. (2002).
This wide-field survey revealed the presence of
stellar members of UMi beyond the previous measured tidal radius, indicating
the existence of a tidal extension in this galaxy. This tidal extension could
be spread out well beyond the area covered in our survey ($R>80\arcmin$), as
suggested by the presence of a ``break'' to a shallower slope observed in
its density profile (see figure 3 in Mart\'\i nez-Delgado et al. 2001).

In this paper,  UMi's observed surface brightness profile is compared
with a theoretical model of tidally disrupted
dSph satellites. The tidal tail is assumed to be formed by the
tidal forces of the Milky Way potential, which are important at
the galactocentric distance of UMi. The tidal forces produce
deformations in the satellite structure and the disruption
of the dwarf. The limit of the bound material, that is, the tidal radius,
is determined by the equilibrium between the satellite potential well and
the Milky Way halo potential well. This equilibrium can also  translate
into a relation between the satellite average density, $\bar{\rho_{\rm S}}(<r_t)$, and
the local and averaged densities of the Milky Way halo, $\rho_{\rm H}(R)$
and $\bar{\rho_{\rm H}}(<R)$ respectively. Details of the tidal limit calculations
are given in G\'omez-Flechoso \& Dom\'{\i}nguez-Tenreiro (2001).
The unbound material, that forms the tidal tail and is placed beyond the
tidal radius, is diffused into the Milky Way halo. As
explained in Johnston, Sigurdsson, \& Hernsquist (1999), the surface density of the tidal
tail  approximately follows a power-law profile in the regions close to
the dwarf.
Obviously, the exponent of this power law
depends on the amount of extra-tidal material close to the dwarf and
has a steeper density profile when the amount of stripped material is
low and vice versa. This amount of stripped material correlates with
the strength of the Milky Way tidal force and therefore  varies along the
orbit, since the satellite travels through regions with different tidal fields.
If the amount of stripped material is large,
the tidal tail will have an almost constant mass density region close to
the satellite (one example of that is the UMi tidal extension) with
a tidal tail mass density similar to the satellite mass density at the tidal radius.
In this case, the tail mass density correlates with the
Milky Way potential at the position of the dwarf, since the satellite
mass density at the tidal radius is proportional to the Milky Way potential,
as we have explained above.
So, for a given halo potential, shape of the satellite density profile, and
 position of the satellite in the halo, the mass density of the tidal tail
in the region close to the tidal radius is fixed independently of the
satellite mass. In Figure 1a the projected mass density profiles of
three evolved satellite models have been plotted.
These satellites have different initial-mass models and  have formed
a tidal tail after a few orbits. As  can be seen in this figure, the
central-region of the projected mass density of the satellites differs
by up to a factor ten when
comparing the different models; however, the tail mass density is
almost the same independently of the models. More details of these
models are given in Section \ref{calibration}.

Using these results, if we know the halo potential field we can estimate the
mass density of the satellite galaxy at the tidal radius and therefore
the tidal tail mass density.
Once the mass density profile of the satellite and its tail are estimated,
the $M/L$ ratio of the dwarf can be calculated by fixing this $M/L$ ratio to
reproduce the observed central luminous surface brightness of the satellite.
Given two satellites with different mass densities and with the same central
luminous surface brightness, the highest mass density satellite will have a
larger $M/L$ ratio than the low mass density one.
Following this relation, if the two satellites with the
same central luminous surface brightness are placed at the same point in
the main galaxy potential, as the mass density of the tidal tail is the same
for both satellite, the tail surface brightness of the denser satellite
will be fainter than that of the low mass density satellite (assuming that a tidal tail
has the same $M/L$ ratio as its satellite).
In Fig. 1b, the three satellite models of Figure 1a have been plotted assuming that
all of them have a central surface brightness of 26 mag arcsec$^{-2}$. As
can be seen in this figure, the denser model needs to have a larger $M/L$ ratio
to reproduce the same central surface brightness as the other models, and
consequently its tidal tail is fainter. Assuming the same central luminous
surface brightness for the satellite models, the differences in the central
mass density  translate into differences in the tail surface brightness.
Only one satellite model will be able to reproduce, at the same time, the central
luminous surface brightness and the tail luminous surface brightness of an
observed satellite dwarf galaxy
at a given position in the halo. Using this method, we can estimate the
$M/L$ ratio of the dwarf independently of its internal kinematics.

\section{CALIBRATION OF THE TIDAL TAIL MODEL}
\label{calibration}

As has been explained in the previous section, the mass density of
a satellite tidal tail is independent of the satellite mass (given
the shape of the satellite density profile and its position in the orbit).
Using the observed surface brightness profile of the satellite, we can
select the model that fits the luminous surface brightness of the satellite + tail
system, and therefore the mass of the best fit model is a good
estimate of the mass of the observed satellite galaxy. Consequently,
the $M/L$ ratio of the satellite dwarf can be calculated.
Before using this method to estimate the mass of the dwarf satellites, we
have to calibrate the sensitivity of the method to small differences in
 satellite mass (in other words, to study whether satellite models with
different masses can reproduce the same satellite + tail density profile or not),
to differences in the shape of the main galaxy potential and to variations
of the dwarf orbit. In this way, we will know the accuracy of the calculations.

\subsection{Effects of  satellite total mass on  tidal tail formation}
\label{satmass}

To analyze how the mass content of a satellite is related to the
tidal tail surface brightness, we have selected three satellite
models with a King--Michie profile with the same dimensionless central
potential, $W_o=4$, and the same core radius, $r_o=0.3$ kpc
(which corresponds to a tidal radius $r_{\rm c}=2.1$ kpc),
but different total mass. The three total masses
are  $0.4\times 10^7, 1.6\times 10^7$, and $4\times 10^7$ $M_{\odot}$.
These satellites have been placed in an orbit similar to that of UMi
(Schweitzer, Cudworth, \& Majewski 1997).
In these models, we have selected an analytical logarithmic potential
to roughly reproduce the Milky Way potential.\footnote{In this section we are only
interested in analyzing the effects of the satellite mass content in the
tidal tail formation and
not in reproducing the details of the  morphological evolution of UMi.} The shape
of this logarithmic potential is
$\Phi_{\rm H} = v^2 \log(R^2+z^2/h^2+a^2)$ where $v=170$ km s$^{-1}$, $a=19.9$ kpc
and $h=1.0$.
For this potential, the apocenter and pericenter
of the orbit of the models are approximately 80 and 20 kpc, respectively.

In Figure 1b, we have plotted the surface brightness of the three
satellite models with different initial total masses.
The initial masses of the models do not give information about their masses
at the moment of the figure snapshot, since the
satellites have undergone tidal disruption. In order to better
compare the models, the masses inside 1 kpc of each model are also listed
in the figure.
The three models have been
calibrated in luminosity to have the same central surface brightness
($\mu_{v,o}=26$ mag arcsec$^{-1}$).
The calibration can be translated in terms of different $M/L$ ratios, as
as  explained in  Section \ref{observations}.
This figure shows that the three brightness profiles have the same
slope in the central regions. However, the luminous surface brightness of the tidal tails
are closely related to the mass content. The lowest-mass satellite, with
$M/L=3.5$, develops  quite a luminous tidal tail. In fact, this satellite only
survives a few orbits in the halo potential. In
 contrast, the highest-mass satellite forms a low brightness tidal tail
($\mu_{v,\mbox{tail}}\sim 33.5$ mag arcsec$^{-1}$) and  has a larger tidal
radius ($r_{\rm t}\sim 1.5$ kpc). The tidal tail of this massive model ($M/L=50$)
will not be observationally detected, assuming a typical satellite central
surface brightness. Finally, the model with a moderate $M/L$ ratio ($M/L=13$)
develops a tidal tail that could be detected with the present observational
resolution, as its tidal tail surface brightness is $\mu_{v,\mbox{tail}}\sim 32$
mag arcsec$^{-1}$.
In the same figure, the observational data for the surface brightness of UMi
are plotted (black dots) in order to compare the models properly.

Summarizing these results, the effect of  satellite mass
content on  tidal tail formation is quite important, as a variation of
a factor 10 in mass produces a variation of a factor of 15 in the
$M/L$ ratio, assuming the same central surface brightness.

\subsection{Effects of  halo potential oblateness of the main galaxy on
 tidal tail formation}
\label{haloshape}

The tidal tail surface brightness of a satellite in a given orbit is related not
only to its mass content but also with the shape of the potential of the main
galaxy. For this reason, we have estimated the importance of the main
galaxy oblateness in  satellite disruption.
We have modeled the halo of the primary galaxy using the logarithmic
potential of the previous section with the same $v$ and $a$ parameters, but
with three different oblateness values, $h=1.0, 0.8$, and $0.6$ (oblateness values smaller
than $0.6$ in the potential are  unrealistic, since they cannot be
reproduced with any mass distribution). The satellite model is a King--Michie
model with initial total mass $M_{\rm ini}=2\times 10^7\ M_{\odot}$,
core radius $r_o=0.3$ kpc and dimensionless central potential $W_o=4$. The
satellite orbit has been fixed on the assumption that it has the energy needed to reproduce
the present position and velocity of UMi.

Figure 1c shows the surface brightness of the
satellite after a few orbits for the three halo potential models. The
snapshots used in this plot represent a satellite with the same
position and velocity as UMi.
This figure shows that it is possible to reproduce the same satellite
and tidal tail profiles using a different halo potential; however, the final
satellite mass content varies.
The formation of dense tidal tails is more efficient
in oblate potentials (for orbits such as that of
 UMi); therefore, the satellite needs
fewer orbits to develop an observable tidal tail. Therefore, for the same
initial satellite model, the larger oblateness of the halo is, the
higher satellite mass content is necessary to reproduce the same
satellite+tail surface brightness profile and, consequently, the larger
$M/L$ will be.

It is remarkable that the effect of the halo potential shape on the
determination of the mass--luminosity ratio is fainter than the effect of
the initial satellite mass. It can be seen that there is only a factor
of two in the $M/L$ ratio for the whole range of  realistic values of
the potential oblateness
(that is, $h$ between 1.0 and 0.6). However, a variation of a factor of 10
in the satellite mass produces a variation of a factor of 15 in the $M/L$
ratio. This means that the potential shape only introduces
a small indeterminacy in the satellite mass calculation.

\subsection{Effects of  small variations in the satellite orbit on
tidal tail formation}

Variations in the satellite velocity within the observational errors
produce differences in the dwarf orbit and, consequently, in its apocenter
and pericenter. As the tidal stripping of  satellites depends on the orbit
(the smaller pericenter of the orbit, the larger the tidal stripping),
small differences in the orbit can affect the tidal tail formation.
In order to calibrate this effect, we have placed the same satellite model in three
different orbits. One of these fits  UMi's velocity and position at the end of the
simulation, while the other two orbits differ by $\pm$10\% in velocity from the
previous one.
The satellite model corresponds to a King--Michie model
with core radius $r_o=0.3$ kpc, initial total mass $M_{\rm ini}=1.6\times 10^7\
M_{\odot}$, and dimensionless central potential $W_o=4.0$.
The halo potential of the main galaxy is the spherical logarithmic potential
described in Section \ref{satmass}.
In Figure 1d, the surface brightness profiles of the models are
shown. These three profiles, corresponding to the three orbits,
are quite similar, only small differences in the
$M/L$ ratio can be observed. The satellite in the highest-velocity orbit
shows the largest $M/L$ ratio and vice versa, however, the $M/L$ ratio
difference between models are of the same order as their difference
in orbital velocity.

The error in the $M/L$ ratio calculation due to the observational
errors of the satellite velocity is of the same order of magnitude as these
observational errors.
For observational errors of 10--20\%, the satellite mass
content is more important in the $M/L$ determination than the effects
introduced by small variations in the satellite orbit.

\section{THE $M/L$ RATIO OF THE URSA MINOR dSph}

UMi shows observational features that reveal
the existence of large tidal forces. Therefore, the simplify
virial equilibrium theorem assumption is not justified for measuring the $M/L$ ratio,
but we should include all the variables in a more generalized expression.
Instead of doing this, we have introduced a
new method of mass calculation that does not involve the kinematics of the dwarf.
In this section, we have analyzed the $M/L$ ratio of the UMi dSph
using the method described in Sections \ref{observations} and \ref{calibration}.

The Milky Way halo density is one of the parameters that determines the
tidal tail density. Therefore assuming a realistic value for it is very
important. The halo density can be deduced from the shape of the
Milky Way halo potential.
Such a potential can be inferred from the dynamics of stars of the Milky Way and the
orbits of the dwarf satellite galaxies. Observational data of the tidal stream
of the Sagittarius dwarf galaxy are therefore very valuable. The shape of the
Sagittarius orbit is traced with the tidal debris of this satellite in
a strip $100^\circ$ long (see Mart\'{\i}nez-Delgado et al. 2002 for details).
We have used these observational data to fix the parameters of the
Milky Way model. This is a quite realistic three-component model
(halo + disk + bulge) that is described in detail in Fux (1997).
The parameters of the final numerical model that
reproduce the Milky Way potential are described in G\'omez-Flechoso et al. (1999).
Other density distributions of the Milky Way can also reproduce a similar
potential well,
compatible with the Sagittarius orbit. It is important to remark
that the physical quantity that determines the satellite orbits is the
potential of the main galaxy, not its density distribution, and that, therefore,
density distributions of the Milky Way halo with different shape and oblateness
could reproduce the same results if they had similar potential wells.
However, we recall that small variations in
the shape of the main galaxy potential do not significantly change  the
results of the $M/L$ ratio of the analyzed satellite, as
shown in Section \ref{haloshape}.

The UMi satellite has been represented by an $N$-body model
orbiting in this Milky Way potential. The orbit is consistent
with the observational proper motion of UMi (Schweitzer et al. 1997).
We assume a
UMi model of one component given by a King--Michie model and therefore the
observed luminosity profile must be
reproduced by the model density profile. The parameters of the
King--Michie model that reproduces more accurately the shape of the density
profile of UMi and its tidal tail are a core radius $r_o=0.3$ kpc and  a
dimensionless central potential $W_o=4.0$. However, the total mass
of the model is still a free parameter that we calculated reproducing
the central brightness of the dwarf and the tidal tail
brightness at the same time (as described previously). The
best fit is obtained with a model of total initial mass
$M_{\rm ini}=4\times 10^7\ M_{\odot}$. The surface brightness profile of the
UMi model and its tidal tail after roughly seven orbits is plotted in
Figure 2 (dashed line). The observational UMi profile has also been
represented in the same Figure (dots). As  can be seen, the model
provides a good fit of the observational data. We have assumed
a mass-to-luminosity ratio $M/L=12$ to adjust the central surface
brightness of the model to that observed. In this snapshot,
the total mass in the inner kiloparsec of the model is $0.79\times 10^7\
M_{\odot}$.

The value of the $M/L$ ratio obtained using this new method
of mass calculations is different from the matter content derived from
the virial theorem ($(M/L)_{\rm virial}=60$; Mateo 1998). However, this
lower value is very consistent with the new   mass-to-light
ratio estimate obtained by Palma et al. (2002), who reduce the mass-to-light ratio
 to $M/L\sim 16$, using a new estimate of the UMi total luminosity
(which is  2.7 times greater than the previous values) and  considering the
(possible) effects of anisotropic velocity dispersions (Hodge \& Michie 1969).

The observed velocity dispersion obtained for UMi is 7.6 km s$^{-1}$
(Armandroff, Olszewski, \& Pryor 1995). In our UMi model, the velocity dispersion
at the center is 4 km s$^{-1}$ and presents a rising profile that reaches
8 km s$^{-1}$ on the outskirts of the dwarf (Mart\'{\i}nez-Delgado,
G\'omez-Flechoso, Alonso-Garc\'{\i}a \& Aparicio 2003, in preparation).
However, the observed velocity dispersion includes the
effects of the substructures of the main body of UMi, and
that could increase the velocity dispersion.
This substructure cannot be reproduced with a numerical model because
we do not have sufficient resolution  to form condensations inside UMi and,
therefore, the velocity dispersion of the UMi models is not expected to fit
that observed. On the other hand, if UMi is assumed to be in virial equilibrium,
the mass-to-luminosity ratio obtained from its observational velocity dispersion
($(M/L)_{\rm virial}=60$) will produce a very low surface brightness tidal
tail, which could not be observationally detected (see Figure 1b).
Furthermore, the signatures of tidal disruption observed in UMi (internal
substructure and tidal extension) make doubtful the existence of  simple virial
equilibrium in the internal region of UMi. In these dynamical conditions, our method
seems to be more reasonable for the satellite mass calculation.
However, new models with a higher resolution reproducing the
internal dynamics of the UMi dSph and new observations of the UMi velocity dispersion
profile are needed in order to understand the dynamical state of the dwarf.

\section{CONCLUSIONS}

We have developed a new method of estimating the mass of  satellite
galaxies   subjected to tidal forces. This method does not involve
suppositions about the internal dynamics of the satellite and can therefore
 be used for satellites that are out of virial equilibrium.

The main results of this model are that massive and dense satellites form low
brightness tidal tails (for a given satellite central surface brightness),
and that  low-density satellites undergo strong tidal forces that produce
comparatively bright tidal tails. Once the satellite central surface brightness is known,
the main parameters that determine the tail brightness
in the region close to the satellite tidal radius are the density profile and the
mass content of the bound part of the satellite.

The shape of the primary galaxy potential can introduce small uncertainties
in the satellite mass estimates that are no larger than a factor of
two over the whole physical range of oblateness. The larger the halo oblateness is, the
denser the tidal tail that is formed.
The observational errors of the satellite orbital velocity can also introduce an
uncertainty in the $M/L$ ratio calculation, but this uncertainty is no larger than
the errors in the satellite velocity.

Obviously, the tidal tail brightness depends on the satellite's position in its
orbit, since tail mass density depends on the halo potential at the
satellite's position. As the satellite travels along its orbit, it goes through different
density regions and  produces variations in the tidal tail brightness. So it is very
important to know the satellite's position in its orbit in order to estimate its mass
content using this method. A satellite tidal tail can be observed when
the satellite is in a given region of its orbit, but it can be too faint to be detected in other regions of the orbit.

Finally, we have applied the new method for the mass calculations  to UMi.
The results shows $M/L \sim 12$, in order to reproduce its luminosity profile and tidal tail
brightness. The tidal disruption features observed in UMi suggest that this
dwarf is not in virial equilibrium. The observational data of the velocity dispersion
profile could confirm this if they show the same rising profile as that of the model.

\clearpage

\begin{figure}
\plotone{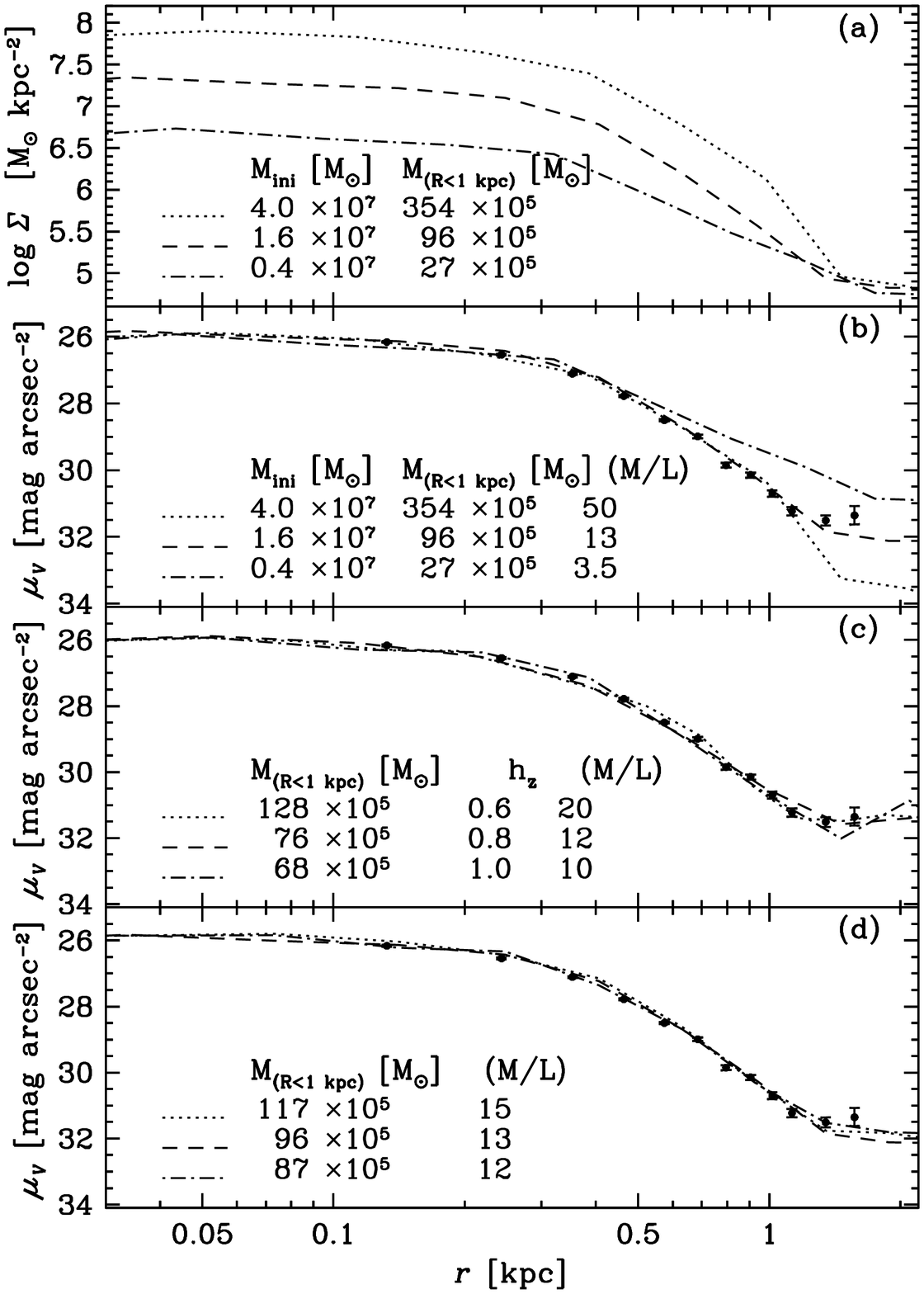}
\caption{(a) Satellite surface density for three
initial satellite masses. The three profiles have the same slope in the
central region. (b) Luminosity profiles for the three satellite models
assuming an $M/L$ ratio in order to obtain a central surface brightness of 26 mag arcsec$^{-2}$.
To compare the effects in a realistic way we have represented
the UMi profile with the error bars in the same plot (dots). (c) Surface brightness profile
for the same initial satellite model but different oblatenesses of the halo
potential. The UMi profile is also plotted (dots with the error bars).
(d) Surface brightness profile for a satellite model placed
in three orbits that have the same satellite position at the final time of the
simulation but differ by $10\%$ in the satellite velocity. Black dots correspond
to the UMi profile.  \label{fig1}}
\end{figure}

\begin{figure}
\plotone{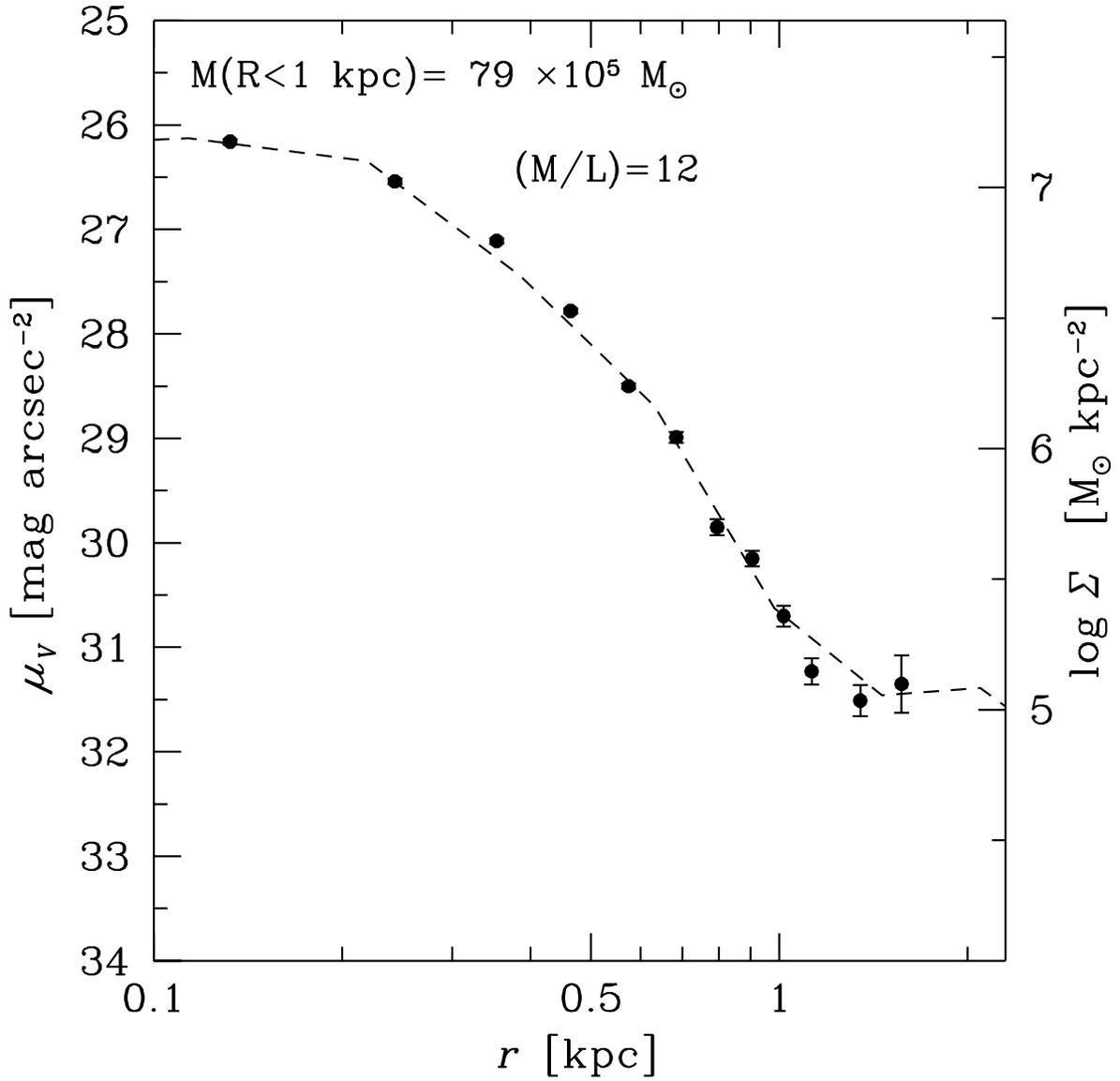}
\caption{Surface brightness profile of the best UMi model (dash line),
an $M/L=12$ has been assumed. The observational data with the error bars are also plotted
(dots). \label{fig2}}
\end{figure}

\end{document}